\documentstyle[amssymb,aps,prl,twocolumn]{revtex}
\begin{document}
\draft

\twocolumn[\hsize\textwidth\columnwidth\hsize\csname@twocolumnfalse\endcsname
\title{Spin Dynamics in the Magnetic Chain Arrays of Sr$_{14}$Cu$_{24}$O$_{41}$: a Neutron
Inelastic Scattering Investigation}

\author{L.P. Regnault$^{a}$, J.P. Boucher$^{b}$, H. Moudden$^{c}$, J.E. Lorenzo$^{d}$, A. Hiess$^{e}$,\\
 U. Ammerahl$^{f}$, G. Dhalenne$^{f}$ and A. Revcolevschi$^{f}$}

\address{$^{a}$D\'{e}partement de Recherche Fondamentale sur la Mati\`{e}re
Condens\'{e}e, SPSMS, Laboratoire de Magn\'{e}tisme et de Diffraction
Neutronique, CENG, F-38054 Grenoble cedex 9, France.\\
$^{b}$Laboratoire de Spectrom\'{e}trie Physique, Universit\'{e} J. Fourier
Grenoble I, BP 87, F-38402 Saint Martin d'H\`{e}res cedex, France.\\
$^{c}$Laboratoire L\'{e}on Brillouin, Centre d'Etudes de Saclay, F-91191
Gif sur Yvette cedex, France.\\
$^{d}$Laboratoire de Cristallographie, CNRS, BP 166, F-38042 Grenoble cedex
9, France.\\
$^{e}$Institut Laue Langevin, BP 156, F-38402 Grenoble cedex 9, France.\\
$^{f}$Laboratoire de Chimie des Solides, Universit\'{e} Paris-Sud, Bat 414,
F-91405 Orsay, France\\}
\date{29 June 1998}

\maketitle

\begin{abstract}
Below T$\approx $150 K, the spin arrangement in the chain arrays
of Sr$_{14}$C$_{24}$O$_{41}$ is shown to develop in two dimensions (2D). 
Both the correlations and the dispersion of the observed elementary excitations
agree well with a model of interacting antiferromagnetic dimers. Along the chains, 
the intra- and inter-dimer distances are equal to 2 and $\approx $3 times the 
distance $(c)$ between neighboring Cu ions. While the intra-dimer coupling is $J\approx 10$
meV, the inter-dimer couplings along and between the chains are of comparable strength, 
$J_{\parallel }\approx -1.1$ meV and $J_{\perp }\approx 1.7$ meV, respectively. 
This remarkable 2D arrangement satisfies the formal Cu valence of the undoped compound. 
Our data suggest also that it is associated with a relative sliding of one chain with 
respect to the next one, which, as T decreases, develops in the chain direction. 
A qualitative analysis shows that nearest inter-dimer spin correlations are ferromagnetic, 
which, in such a 2D structure, could well result from frustration effects.

\end{abstract}

\

] \narrowtext

An increasing interest is presently devoted to the compound Sr$_{14}$Cu$%
_{24} $O$_{41}$ as superconductivity can be obtained in Ca doped materials
under pressure \cite{Uehara96}. This compound is also a remarkable spin
system, which is made of two distinct magnetic structures\cite{Carron88}.
One structure consists of an array of quantum ($s=1/2$) two-leg Cu$_{2}$O$%
_{3}$ ladders and the other of an array of CuO$_{2}$ quantum spin chains.
These two mixed spin subsystems are expected to play a crucial role in the
occurrence of superconductivity. An important and first question concerns
the charge transfer process which takes place upon doping.\ Due to the
formal Cu valence ($+2.5$), one expects holes to be already present in
stoichiometric compounds. Recently, it has been proposed that, in such
undoped materials, most of the holes are located in the chains\cite{Kato96}.
This should result in a specific spin distribution in the chains. It is the
purpose of the present neutron inelastic scattering (NIS) investigation to
explore the actual spin arrangement in the chain arrays of the undoped
compound. A good knowledge of the initial hole distribution is definitely
required before any investigation of the hole-doping process be further
developed. A few neutron investigations on undoped Sr$_{14}$Cu$_{24}$O$_{41}$%
have been previously performed. The first measurements were carried out on a
powder sample\cite{Eccleston96}. The presence of magnetic excitations in the
vicinity of 11 meV was confirmed, in agreement with susceptibility
measurements\cite{Matsuda96}. They were attributed to the formation of
antiferromagnetic (AF) dimers in the chains. The first measurements on a
single crystal have revealed the presence of several excitation branches in
the same energy range\cite{Matsuda(a)96}. Their dispersion, which was
determined in a limited part of the Brillouin zone, provides evidence of
interactions between the AF dimers in the chain direction. It was then
proposed that the dimers are formed between spins separated by ``$2$ and $4$%
'' times the distance ($c$) between the nearest-neighbor Cu ions in the
chains. In these measurements, however, the chosen scattering plane
prevented any direct exploration of the chain arrays (fig. 1a). More
recently, using a set of eight single crystals, new measurements were
performed\cite{Eccleston98}. The dispersion along the chains was determined
with an improved accuracy. The data were shown to compare well to a
one-dimensional (1D) model of interacting AF dimers (also discussed in\cite
{Takigawa98}), where the intra and inter-dimer distances ($d$ and $d^{\prime
})$, respectively) correspond to $2$ or $3$ times the distance $c$. Due to
averaging effects inherent to the experimental procedure, again no
information relative to any transverse directions could be obtained. In the
present investigation, a unique single crystal was used and the measurements
were performed on triple-axis spectrometers (IN8 and 1T1 at the Institut
Laue Langevin in Grenoble and at the Laboratoire L\'{e}on Brillouin in
Saclay, respectively). The orientation of the crystal was chosen in such a
way that a rather exhaustive investigation of the chain arrays could be
developed.

Our Sr$_{14}$Cu$_{24}$O$_{41}$ sample was cut from a crystal several cm long
grown by the travelling solvent zone method under a pressure of $3$ bar
oxygen atmosphere\cite{Revco97}. Its volume was about $5\times 5\times 30$ mm%
$^{3}$ with the {\bf c} axis in the longer dimension. The crystallographic
structure of this material can be described with two distinct unitary
subcells, one referring to the ladders, the other to the CuO$_{2}$ magnetic
chains. For the latter subcell - to be considered hereafter - the space
group and the lattice constants are at room temperature: A{\it mma}, $%
a=11.456$ $A$, $b=13.361$ $A$, $c=2.749$ $A$,\cite{Carron88}. A sketch of
the chain arrays in the ({\bf a},{\bf c}) plane is shown in fig. 1a. In the 
{\bf a }direction, one observes a doubling of the unit cell (the lattice
parameter is twice the distance between chains) which is due to the relative
shift (in the {\bf c} direction) of one chain with respect to the next one%
\cite{Matsuda97,Cox98}. This relative displacement is small at room
temperature, $\eta $ $\approx 0.17c$\cite{Matsuda97}. Most of our neutron
measurements were performed with the scattering wavevector ${\bf Q}$ lying
in the ({\bf a},{\bf c}) plane. For inelastic scatterings, a horizontal
collimation of 50'-50'-open-open was used, with a final wavevector ${\bf k}%
_{f}=2.662$ $A^{-1}$. Pyrolytic graphite single crystals were mounted as
monochromator and analyser, and a pyrolytic graphite filter after the sample
was used to eliminate higher-order flux contamination. Most of the energy
scans reported here were performed in the range $2-20$ meV with an energy
resolution of the order of $1$ meV (full width half maximum). The
temperature could be varied from $1.5$ to $300$\ K. In general, as shown in
fig. 2a, two (and only two\cite{Matsuda(a)96}) well-defined magnetic
excitations are observed in the ({\bf a},{\bf c}) plane at low temperature.
For a few wavevectors, however, only a single excitation is detected. It is
the case for the wavevectors ${\bf Q}=(2.5,0,Q_{L})$, where the unique line
results from the superposition of two excitations (cf. fig. 2a which shows
the single peak at $Q_{L}=0.25$). More surprisingly, it is also the case for
the specific wavevectors $(Q_{H},0,0.65)$ (see fig. 2b) where an
``apparent'' extinction of one of the two modes occurs.\ Figs. 3a and 3b
display the dispersions of the modes observed along the {\bf a} and {\bf c}
directions, respectively. The former figure reports the data obtained at $%
Q_{H}=2$ and $Q_{H}=3$. For these two values of $Q_{H}$, the data superpose
very well yielding an excellent definition of the periodicity of the
propagation along {\bf c}. The latter figure displays a new and important
result. Dispersive modes are also observed in the {\bf a} direction. This
result establishes unambiguously that a magnetic coupling does exist between
the chains. As there are $2$ non-equivalent neighboring chains in the {\bf a}
direction (they form a two-Bravais lattice ), one is led to analyse the two
observed branches (E' and E'' in fig. 3b) as resulting from an unique
excitation, which should be twice degenerated without such interchain
interactions. The degeneracy is lifted by the interchain couplings and two
propagative dispersions are observed. One branch is to be considered as the
``image'' of the other, shifted by one reciprocal lattice unit (r.l.u.) $%
2\pi /a$ (see discussion below). In contrast to what is proposed in\cite
{Matsuda(a)96}, the two branches do not result from the presence of possible
magnetic anisotropies. On the contrary, the spin system describing the
chains is to be considered as isotropic and, accordingly, the two branches
are expected to be associated with a $S=1$ magnetic state. Under the effect
of a magnetic field, they undergo an identical Zeeman splitting giving rise
to a unique electron spin resonance (ESR) line as observed experimentally%
\cite{Matsuda96}. With such an analysis, no contradiction is found between
neutron and ESR data. Another striking result revealed by our measurements
concerns the integrated intensity $I_{Q}$ of the observed excitations. The
values of $I_{Q}$ measured in the chain direction are reported in fig. 4.
Several features are worth being noted. In particular, extinction is seen to
occur at specific values of $Q_{L}$: 0, 1/2 and 1. For instance, for $%
Q_{H}=3 $ (fig. 4a), a double peak structure is seen in the first half of
the Brillouin zone and a single peak structure in the second half. All these
features must be taken into account in any analysis of the magnetic
properties of the chains.

As a starting point of our analysis, we consider the 1D model of interacting
AF dimers previously proposed\cite{Matsuda96,Takigawa98}. In such a case, we
may refer to the description of interacting excitons given in\cite{Harris73}%
.\ When the intra-dimer coupling $J$ is much larger than the inter-dimer one 
$J_{\parallel }$, the energy dispersion along the chain is given by\newline
$E/J\approx (1-\alpha _{\parallel }^{2}/4)-\alpha _{\parallel }(1+\alpha
_{\parallel })\cos (m2\pi Q_{L})+(\alpha _{\parallel }^{2}/4)\cos (m4\pi
Q_{L})\approx 1-\alpha _{\parallel }\cos (m2\pi Q_{L})$ \newline
with $\alpha _{\parallel }=J_{\parallel }/2J.$ In that expression, $Q_{H}$
is expressed in the r.l.u. $2\pi /c$, the inter-dimer distance is fixed to $%
d=2c${\rm \ }(see discussion below) and the parameter $m$ [$=(d+d^{\prime
})/2$] allows a determination of the inter-dimer distance $d^{\prime }$
relatively to $d$. A fitting procedure where $J$, $\alpha _{\parallel }$ and 
$m$ are adjustable parameters is applied successively to the two sets of
data in fig. 3. Within the experimental accuracy, identical values are
obtained for $\alpha _{\parallel }$ and $m$: $\alpha _{\parallel }=-0.065\pm
0.007$ and $m=4.95\pm 0.3$. Since $\alpha _{\parallel }<0$, the inter-dimer
coupling $J_{\parallel }$ is deduced to be ferromagnetic (F) in that
description. The value obtained for $m$ suggests the inter-dimer distance $%
d^{\prime }$ to be practically equal to $d^{\prime }=3c$ \cite
{Eccleston98,Takigawa98}. In that 1D description, different values are
obtained for $J$ : $J_{1}\approx 12.0$ meV and $J_{2}\approx 10.3$ meV for
the upper and lower branches, respectively. As discussed above, however, the
energy difference between the two branches results from the small dispersion
observed in the transverse direction. Extrapolating the 1D description used
above to the presence of small interchain couplings ($J_{\perp }\ll J$), we
can write the dispersion observed in the {\bf a} direction as $E/J\approx
1-\alpha _{\perp }$cos$(\pi Q_{H}+\varphi )$ with $\alpha _{\perp }=J/2$ and
where the phase factor $\varphi $ ($=0$ or $\pi $) accounts for the two
transverse dispersive branches E' and E''. Finally, after fitting the data
in fig. 3b (the full and dashed lines with $\mid \alpha _{\perp }\mid
=0.075\pm 0.007$), one obtains the following set of values: $J\approx 10$
meV, $J_{\parallel }\approx -1.1$ meV and $J_{\perp }\approx 1.7$ meV. It is
established here that the couplings between AF dimers along {\bf c} and
along {\bf a} are of comparable strength.

In a second step of our analysis, we consider the data of the integrated
intensity $I_{Q}$ reported in fig. 4. At low temperature, as higher energy
excitations can be reasonably ignored, the quantity $I_{Q}$ provides a good
evaluation of the ``structure factor'' $S_{Q}=\Sigma _{n}\langle
s_{n}s_{n^{\prime }}\rangle \exp [iQ(n-n^{\prime })]$ where $\langle
s_{n}s_{n^{\prime }}\rangle $ defines the static 2-spin correlation
functions for spins located at sites $n$ and $n^{\prime }$. Since $%
J_{\parallel }$, $J_{\perp }\ll J$, one expects $S_{Q}$ to be dominated by
the intra-dimer correlations. Assuming the spins of a dimer to be located in
a chain at sites $n-1$ and $n+1$, the different intra-dimer correlations are
about the same in absolute values, i.e. $\langle s_{n\pm 1}s_{n\pm 1}\rangle
\approx \mid \langle s_{n\pm 1}s_{n\mp 1}\rangle \mid $. As the excitation
energy ($\approx 11$ meV) is large compared to the dispersion amplitude ($%
\approx 1$ meV), the correlation length $\xi $ of that quantum system is
short ($\xi \ll c,a$). Accordingly, as a first approach, we may limit the
sum $\Sigma _{n}$ to the inter-dimer correlations between nearest neighbors.
Then, assuming (in the chain direction) $\langle s_{n\mp 1}s_{n\pm 4}\rangle
\approx \langle s_{n\pm 1}s_{n\pm 6}\rangle \approx -\langle s_{n\mp
1}s_{n\pm 6}\rangle \approx -\langle s_{n\pm 1}s_{n\pm 4}\rangle $, we
obtain from the correlations along the chains the expression $I_{Q}\approx
[1-\cos (4\pi Q_{L})][1+2\varepsilon _{\parallel }\cos (m2\pi Q_{L})]$ where
the first factor accounts for the intra-dimer correlations and $\varepsilon
_{\parallel }$ is the ratio between the inter- and intra-dimer correlations (%
$\varepsilon _{\parallel }=\langle s_{n-1}s_{n+4}\rangle /\langle
s_{n-1}s_{n-1}\rangle $). The first factor predicts zero intensities at $%
Q_{L}=0,1/2,1$, exactly as it is experimentally observed (see fig. 4). This
explains the value we have chosen above: $d=2c$. In the second factor, the
coefficient $\varepsilon _{\parallel }$ is positive (negative) for AF (F)
inter-dimer correlations. From the analysis of the dispersion curve
presented above, $\varepsilon _{\parallel }$ is expected to be negative and $%
m\approx 4.5$. For $\varepsilon _{\parallel }=-0.2$, one obtains the dashed
line in fig. 4a. Interestingly, the double peak structure observed
experimentally is reproduced qualitatively. In the whole Brillouin zone,
however, the curve remains (almost) symmetric with respect to $Q_{L}=1/2$.
The occurrence of the double peak structure depends directly on the values
of $\varepsilon _{\parallel }$: we evaluate $0.15\lessapprox \mid
\varepsilon _{\parallel }\mid \lessapprox 0.25$. Within the same simplified
description, the effect of spin correlations between dimers located in
neighboring chains can be described in a similar way and one is led to
complete the above expression as (for simplicity, the magnetic form factor
of the Cu$^{2+}$ ions is ignored)\newline
$I_{Q}\approx [1-\cos (4\pi Q_{L})][1+2\varepsilon _{\parallel }\cos (m2\pi
Q_{L})][1+2\varepsilon _{\perp }\cos (\pi Q_{H}+m^{\prime }\pi Q_{L})]$%
\newline
where the last factor takes into account the relative displacement between
neighboring chains: $\eta =m^{\prime }c/2$. Similar to the previous case, $%
\varepsilon _{\perp }$ is the ratio between transverse inter- and
intra-dimer correlations. Depending on whether $Q_{H}$ is even or odd, a
phase factor is seen to be present or not in the cosine function in the last
factor of $I_{Q}$. Considering first the $I_{Q}$ data relative to the
excitation branch E' observed at $Q_{H}=3$ (open dots in fig. 3a), a fitting
procedure, with $\varepsilon _{\perp }$ and $m^{\prime }$ as adjustable
parameters ($\varepsilon _{\parallel }=0.2$ fixed), yields the continuous
curve drawn in fig. 4a, and provides the following evaluations: $\varepsilon
_{\perp }=+0.36\pm 0.06$ and $m^{\prime }=1\pm 0.1$. The non symmetric
behavior observed experimentally is now better reproduced in the whole
Brillouin zone. The positive value obtained for $\varepsilon _{\perp }$
(comparable to $\varepsilon _{\parallel }$, as expected) predicts the
interchain correlations to be AF. Remarkable also is the value obtained for $%
m^{\prime }$ which shows that the relative displacement $\eta $ between
neighboring chains is of the order of $c/2$ at low temperature. In order to
test further these predictions, the same model with all the above parameters
fixed ($\varepsilon _{\parallel }=0.2$, $\varepsilon _{\perp }=+0.36$ and $%
m^{\prime }=1$) is applied to the $I_{Q}$ data relative to the other
excitation branches (figs. 4b and 4c). According to the description proposed
above, the branch E''observed at $Q_{H}=3$ (black dots in fig. 3) can be
viewed as the ``image'' of the excitation branch E' at $Q_{H}=2$. The
corresponding structure factor is therefore to be described by the same
expression as before but without the factor $\pi $ in the last factor of $%
I_{Q}$. One obtains the full curve shown in fig. 4b (there is no additional
adjustment except for an amplitude factor). Again the agreement appears to
be reasonable in the whole Brillouin zone (for comparison, the curve
obtained with the factor $\pi $ is shown as the dashed line). Considering
finally the data obtained at $Q_{H}=2$, the structure factor relative to the
branches E' and E'' (in fig. 3, the open and full squares, respectively) are
also reasonably well explained by the same description: the full and dashed
curves shown in fig. 4c are obtained with the factor $\pi $ included and not
included, respectively.\ The only noticeable discrepancy occurs for the
low-energy data (full squares) in fig. 4c near $Q_{L}=0.65$. Remember (cf.
fig. 2b), it is also the $Q_{L}$ region where an unexplained ``apparent''
extinction of one of the two branches occurs.

The relative displacement $\eta $ can be supposed to vary with temperature.
Such a change in the lattice structure could result in a simultaneous change
in the spin arrangement. This seems to be the case as shown by the data
reported in figs. 5a and 5b. The former figure represents the intensity $%
A_{Q}$ of two nuclear Bragg peaks, observed at ${\bf Q}=(4,0,0.3)$ and ${\bf %
Q}=(3,0,0.3)$. For the chains, these peaks correspond to the same $Q_{L}=0.3$%
, but to different $Q_{H}$ ($=3$ and $4$, respectively). Due to the doubling
of the unit cell in the ${\bf a}$ direction, Bragg peaks for even values of $%
Q_{L}$ are expected not to be strongly affected by a variation of $\eta $.
For odd values of $Q_{H}$, however, $A_{Q}$ should decrease with $\eta $ (it
is equal to zero for $\eta =0$). Figure 5a agrees well with that description
and confirms the picture suggested above, that neighboring chains are slowly
sliding along ${\bf c}$ as $T$ decreases. Simultaneously, changes in the
excitation spectrum are observed (see fig. 5b).\ In that figure, the energy
of the single excitation shown in fig. 2b, is measured at $Q_{H}=3$ and $%
Q_{H}=2$. Due to the transverse dispersion, two different values are
obtained at low temperature. As $T$ increases, however, they are seen to
merge into the same value $\approx 11.3$ meV, showing that the interchain
coupling becomes ineffective above $T\approx 150$K.

The purpose of the analysis sketched above is to draw attention on the very
peculiar spin distribution appearing in the chains of Sr$_{14}$Cu$_{24}$O$%
_{41}$. At low temperature ($T<10$ K), the interactions between the AF
dimers develop in two dimensions (2D). For the parameters determined above ($%
d=2c$, $d^{\prime }\approx 3c$ and $\eta \approx c/2$), different spin
arrangements can be proposed. However, as shown in fig. 1b, there exists one
distribution which corresponds to a complete alternation in the spin
dimerisation. It is remarkable that a similar situation is found in the
dimerised phase of the spin-Peierls compound CuGeO$_{3}$, where a splitting
in two excitation branches has also been recently observed\cite{Lorenzo98}.
For such a symmetric situation, however, if one of the inter-dimer couplings
(or both of them) are assumed to be ferromagnetic, one is inevitably led to
a misfit in the spin correlations, i.e. in the spin arrangement, as
illustrated in fig. 1b. This picture suggests that frustration effects
(typically AF couplings between next neighbor spins) might also play a role,
offering then the possibility of alternative descriptions. Anyhow, to
complete the above analysis, more theoretical and/or numerical studies are
definitely required. The effect of Ca doping is expected to modify the
charge distribution and therefore the spin arrangement in the chain arrays.
For low doping levels (or non-stoichiometric oxygen contents for instance%
\cite{Hiroi96}) the presence of dimers can still be expected, but possibly
with a smaller average inter-dimer distance $d^{\prime }$. This could
explain the periodicity of 2 and 4 reported in\cite{Matsuda(a)96}. For
stronger doping levels, however, more drastic changes should occur.

One of us (JPB) would like to acknowledge J.Y. Henry, T. Ziman, D. Grempel,
F. Mila and G. Bouzerar for numerous and stimulating discussions.

\begin{figure}[tbp]
\caption{a) Schematic representation of a chain array in Sr$_{14}$Cu$_{24}$O$%
_{41}$, with lattice parameters $a$ and $c$. $\eta $ is the relative
displacement between neighboring chains; b) spin-dimerisation in the
alternation model with an illustration of the spin correlations (see text).
The full and open symbols represent Cu and O atoms, respectively}
\end{figure}

\begin{figure}[tbp]
\caption{Examples of chain excitation modes at low temperature ($T<4$ K): a)
energy scans at $(Q_{H},0,0.25)$ and b) at $(Q_{H},0,0.65)$.}
\end{figure}

\begin{figure}[tbp]
\caption{a) dispersion in the chain direction {\bf c} for excitations
observed at $Q_{H}=2$ and $Q_{H}=3$ at low temperature ($T<4$ K); b)
dispersion of the two excitations branches E' and E'' observed in the
transverse direction {\bf a} ($T<4$ K). The lines are fitted curves (see
text). }
\end{figure}

\begin{figure}[tbp]
\caption{Structure factors of the chain excitations as a function of $Q_{L}$%
: a) for the branch E' probed at $Q_{H}=3$; b) for the branch E'' probed at $%
Q_{H}=3$; for the branches E' and E'' (open and full symbols, respectively)
probed at $Q_{H}=2$. The lines are explained in the text.}
\end{figure}

\begin{figure}[tbp]
\caption{a) Intensity of the two nuclear Bragg peaks observed at $(4,0,0.3)$
and $(3,0,0.3)$ as a function of temperature; b) energy of the two
excitation modes observed at $(3,0,0.65)$ and $(2,0,0.65)$ as a function of
temperature. The lines are guides to the eye.}
\end{figure}


\begin{references}
\bibitem{Uehara96}  M. Uehara et al., J. Phys. Soc. Jpn {\bf 65}, 2764
(1996); H. Mayaffre et al. Science, 279, 345 (1998).

\bibitem{Carron88}  E.M. Caron III et al., Mater. Res. Bull. {\bf 23}, 1355
(1988); T. Siegrist et al., id. 23, 1429 (1988).

\bibitem{Kato96}  M. Kato et al., Physica C258, 284 (1996).

\bibitem{Eccleston96}  R.S. Eccleston et al.; Phys. Rev. {\bf B53}, R14721
(1996).

\bibitem{Matsuda96}  M. Matsuda and K. Katsumata, Phys. Rev. {\bf B54},
12201 (1996).

\bibitem{Matsuda(a)96}  M. Matsuda et al., Phys. Rev. {\bf B54}, 12199
(1996).

\bibitem{Matsuda97}  M. Matsuda et al., Phys. Rev. {\bf B56}, 14499 (1997).

\bibitem{Eccleston98}  R.S. Eccleston et al. cond-mat. 9711053.

\bibitem{Cox98}  D.E. Cox et al., Phys. Rev. {\bf B57}, 10750 (1998).

\bibitem{Takigawa98}  M. Takigawa et al., Phys. Rev. {\bf B57}, 1124 (1998).

\bibitem{Revco97}  A. Revcolevschi et al. Physica {\bf C282-287}, 493
(1997); U. Ammerahl et al., J. of Crystal Growth (in press).

\bibitem{Harris73}  Brook Harris, Phys. Rev. {\bf B7}, 3166 (1973).

\bibitem{Lorenzo98}  E. Lorenzo et al. submitted to Phys. Rev. Lett. (1998).

\bibitem{Hiroi96} Z. Hiroi et al., Phys. Rev. {\bf B54}, 15849 (1996).
\end{references}
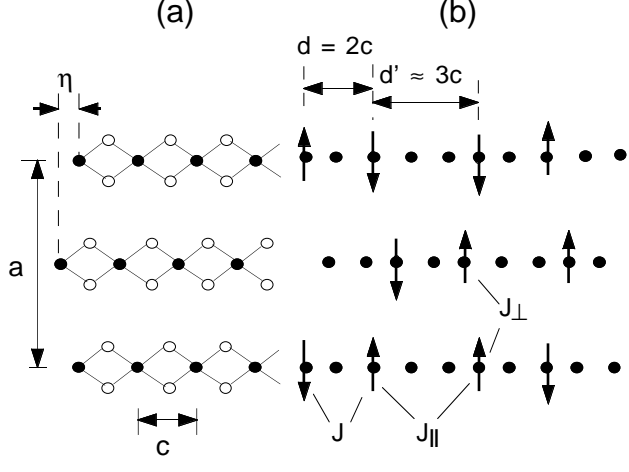
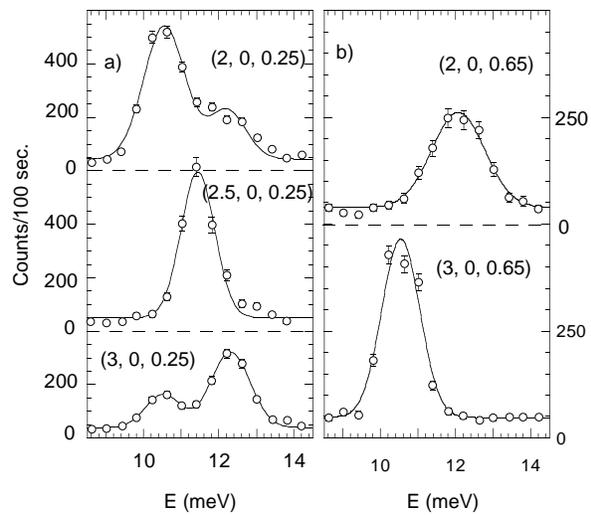
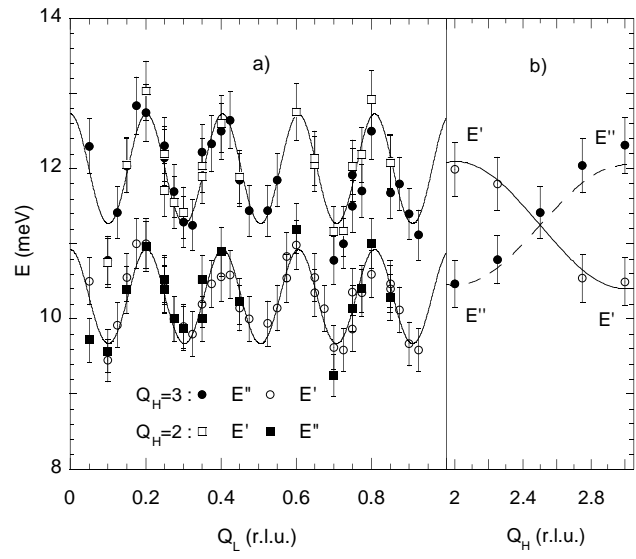
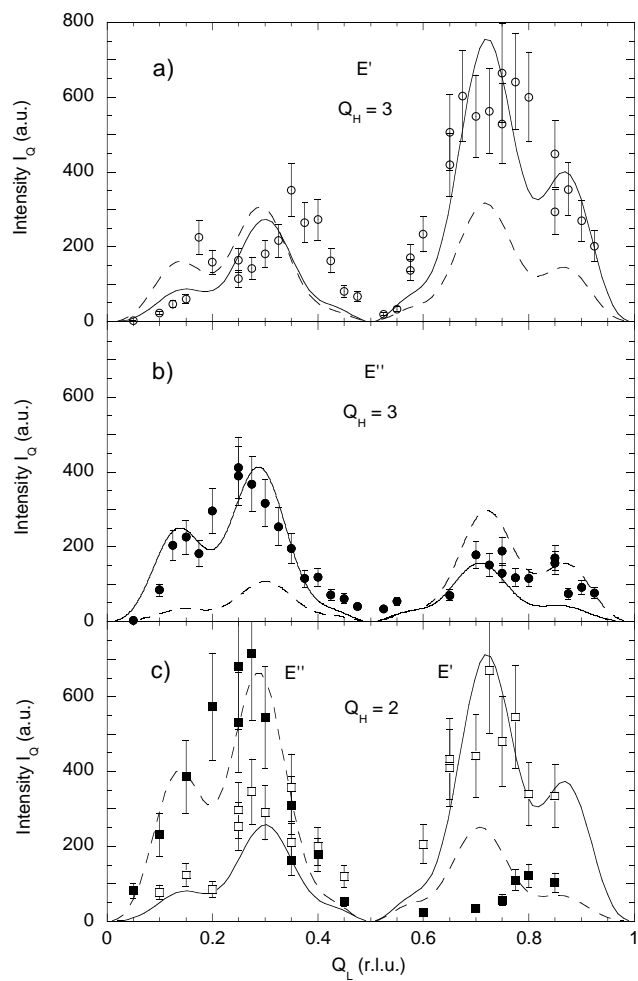
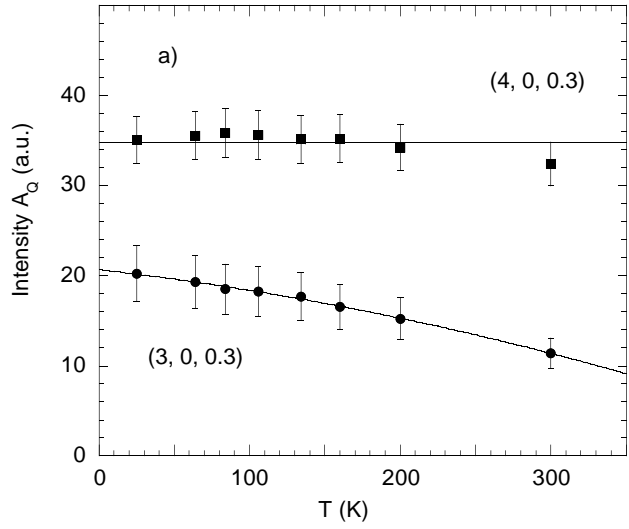
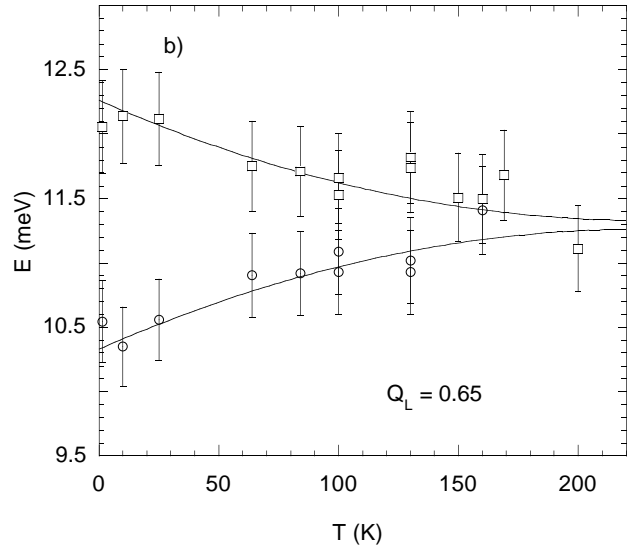
\end{document}